\documentclass[pra,showpacs,twocolumn,superscriptaddress,amsmath,amssymb]{revtex4-1}
 
\usepackage{graphicx}

\usepackage{color}
\usepackage{bm}
\usepackage{amssymb}

\usepackage{hyperref}
\hypersetup{
    colorlinks=true,
    linkcolor=blue,
    citecolor=blue,   
    urlcolor=blue
    }
    
\mathchardef\mhyphen="2D

\begin{document}

\title{Quantum Control of Atom-Ion Charge Exchange via Light-induced Conical Intersections}

\author{Hui Li}
\thanks{Present address: JILA, University of Colorado, Boulder, Colorado 80309, USA}
\address{Department of Physics, Temple University, Philadelphia, Pennsylvania 19122, USA}
\author{Ming Li}
\address{Department of Physics, Temple University, Philadelphia, Pennsylvania 19122, USA}
\author{Alexander Petrov}
\address{Department of Physics, Temple University, Philadelphia, Pennsylvania 19122, USA}
\author{Eite Tiesinga}
\address{Joint Quantum Institute, National Institute of Standards and Technology and University of Maryland, Gaithersburg, Maryland
20899, USA}
\author{Svetlana Kotochigova}
\email{skotoch@temple.edu}
\address{Department of Physics, Temple University, Philadelphia, Pennsylvania 19122, USA}

\date{\today}

\begin{abstract}
Conical intersections are crossing points or lines between two or more adiabatic electronic potential energy surfaces in the multi-dimensional coordinate space of  colliding atoms and molecules. Conical intersections and corresponding non-adiabatic coupling can greatly affect molecular dynamics and chemical properties. In this paper, we predict significant or measurable non-adiabatic effects in an ultracold atom-ion charge-exchange reaction in the presence of laser-induced conical intersections (LICIs).  We investigate the fundamental physics of these LICIs on molecular reactivity under unique  conditions: those of relatively low laser intensity of $10^8$ W/cm$^2$ and ultracold temperatures below 1 mK.  We predict irregular interference effects in the charge-exchange rate coefficients between K and Ca$^+$ as functions of  laser frequency. These irregularities occur in our system due to the presence of two LICIs.  To further elucidate the role of the LICIs on the reaction dynamics, we  compare these rate coefficients with those computed for a system where the CIs have been ``removed''.  In the laser frequency window, where  conical interactions are present, the difference in rate coefficients can be as large as $10^{-9}$ cm$^3$/s.  
\end{abstract}

\maketitle


\section{Introduction}

Theoretical descriptions of many chemical processes are often based on the Born-Oppenheimer (BO) approximation \cite{BO}. This approximation relies on the observation that the motion of electrons and atomic nuclei occurs on different time or energy scales and thus allows the separation of electronic and nuclear degrees of freedom. The resulting model simplification  minimizes computational efforts. 
 
There exist, however, many cases where the coupling between electronic and nuclear motion is non-negligible and  the BO approximation breaks down \cite{Baer2006}. In particular, the potential energy surfaces of electronic states with the same symmetry of polyatomic molecules with three or more atoms can be degenerate at crossing points or curves in the multi-dimensional space of the nuclear coordinates. These crossings are called conical intersections (CIs). Generally, CIs form a seam of $3N-8$ dimensions, where $N$ is the number of atoms in the molecule. At these crossings strong non-adiabatic transitions between electronic states occur. 

Conical intersections between molecular electronic potential surfaces thus greatly affect molecular dynamics and chemical reactivity. Conditions under which conical intersections occur have been extensively reviewed Refs.~\cite{Domcke2004, Domcke2011,Yarkony2016}. Naturally occurring CIs play a crucial role in photochemistry and photobiology as well as optical and superconductor physics \cite{Klessinger1995,Domcke2004,Peleg2007,Leone2008}. Clearly, the location of such CIs and the strength of the related non-adiabatic couplings are inherent properties of the atoms in the molecule and are difficult to manipulate or control.

When molecules are exposed to resonant laser light, however, new features can emerge. Moiseyev {\it et al.}\cite{Moiseyev2008} showed  that  with laser light it is possible to create a conceptually different kind of CI even in diatomic molecules, a so-called light-induced conical intersection (LICI). In contrast to natural conical intersections the characteristics of LICIs are easily modified by the parameters of the laser field. The internuclear positions of  LICIs are determined by the laser frequency and the direction of the laser polarization  $\epsilon$. That is, the angle $\theta$ between a molecular axis and the direction of polarization adds a motional degree of freedom and a controllable CI can appear. 

Initial theoretical descriptions of LICIs  \cite{Moiseyev2008,Moiseyev2011,Sindelka2011,Szidarovszky2018,Szidarovszky2018_1} focused on the diatomic Na$_2$ molecule interacting with  either a retroreflected laser beam creating a spatial standing wave potential for the dimer or with short but strong laser pulses. For both implementations this leads to a degeneracy of ``dressed' electronic states in a two-dimensional space, where one of the coordinates is the separation $R$ between the sodium atoms.
Study of the core excited CO$^*$ molecule in x-ray regimes \cite{Demekhin2011,Demekhin2013} show that LICIs should lead to nonadiabatic transitions between  electronic, vibrational, and rotational degrees of freedom of the diatomic molecule. In addition, with the readily availability of laser-cooled  atoms there is also a growing interest in these concepts applied to ultracold atom-atom collision \cite{Hutson2011,Wuster2011, Toth2019}. The dramatic effect of the light-induced conical intersection on the photodissociation and photofragmentation of the D$_2^+$ molecule was demonstrated in \cite{Halasz2013, Halasz2015, Csehi2016}. Recently,  Csehi {\it et al.}\cite{Csehi2017,Csehi2022}  in a theoretical studies showed  that LICI in diatomics can be created even by quantized radiation field in an optical cavity.  

Light-induced conical intersections have been studied experimentally with ultrafast molecular processes. Kim {\it et al.}\cite{Kim2012}  found that a LICI can  control the isomerization of 1,3-cyclohexadiene,
while Corrales {\it et al.}\cite{Corrales2014}  investigated the transition from a weak- to strong-fields  in the LICI-induced dissociation  of polyatomic methyl iodide. Quantum interference in the dissociation of H$_2^+$ initiated by a LICI was observed by Natan {\it et al.}\cite{Bucksbaum2016} with focussed 30 fs laser pulses with a peak intensity of 2 $\times$ 10$^{13}$ W/cm$^2$. Finally, K{\"u}bel {\it et al.}\cite{Kubel2020} performed detailed investigations of the light-induced molecular potentials in the same molecular system. They demonstrated the presence of distortions in the nonadiabatic potential energy surfaces from  modulations in the angular distribution of the reaction paths.

\begin{figure} [t]
\includegraphics[scale=0.27,trim=0 0 0 0,clip]{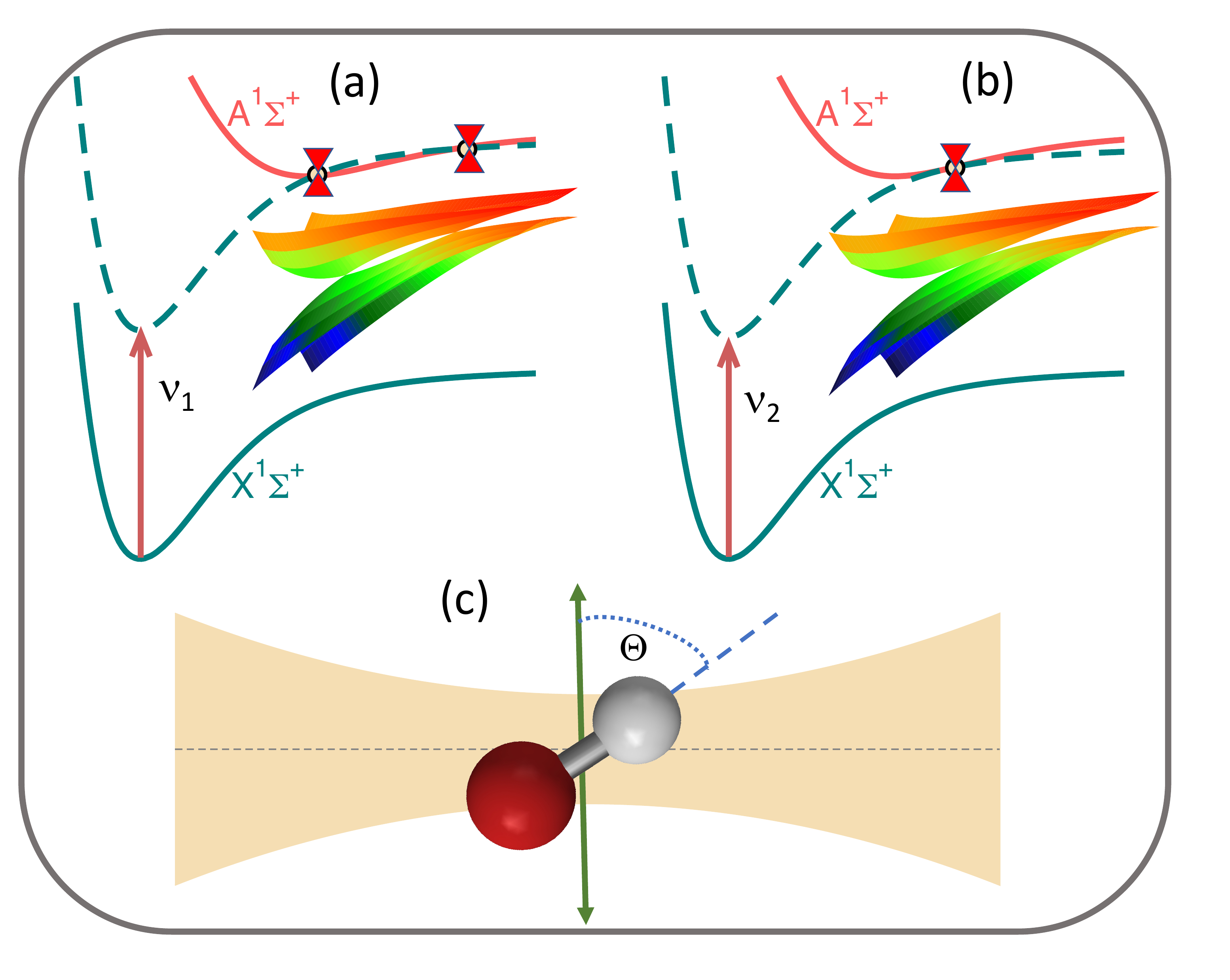}
\caption{Schematic of the LICI-enhanced charge-exchange reaction in ultracold K+Ca$^+$ collisions. 
Panels a) and b) show the photon dressing of the radial diabatic potentials (line drawings) and 
two-dimensional adiabatic potentials of KCa$^+$ as functions of radial separation $R$ and orientation $\theta$  (colored three-dimensional surfaces) with (a) two conical intersections and (b) with one glancing intersection in the $R\mhyphen\theta$ plane, respectively. The laser photon frequencies $\nu_i$ in panels a) and b) are $\nu_1/c$ =  13\,966 cm$^{-1}$ and  $\nu_2/c$ = 13\,889 cm$^{-1}$, respectively. Here, $c$ is the speed of light in vacuum.
Panel c) shows the definition of  $\theta$,  the angle between the inter-atomic axis (blue dashed line) connecting the atom and the ion (red and grey balls) and the direction of the linear laser polarization (green double arrow). The orange region represents the laser intensity profile (not to scale).}
\label{AdiabaticPots}
\end{figure}

In this paper, we investigate the role played by LICIs in the charge-exchange reaction between ground-state $^{40}$Ca$^+$ ions and neutral ground-state $^{39}$K atoms prepared in well-controlled quantum states and at ultracold collision energies.
We  envision two experimental realizations of our ideas. The first 
corresponds to one where ultracold K and Ca$^+$ are stored in overlapping optical dipole and Paul traps, respectively, and the frequency  of a linearly polarized laser is tuned. In a dressed-state picture or with Floquet analysis, dressed molecular states satisfy the conditions required for  LICIs as schematically shown in Fig.~\ref{AdiabaticPots}. 
Two dressed adiabatic molecular states as functions of the atom-atom separation $R$ and the angle between the inter-atomic axis and the laser polarization $\theta$  touch forming one or two LICIs at one or two pairs  $(R,\theta)$. 
The non-trivial dependence of the charge-exchange  rate coefficient on  laser frequency can then reveal the presence of LICIs.
Our simulations will be restricted to this realization.

A second realization, not studied here, to show the presence of LICIs might correspond to the case where  spatially separate clouds of ultracold K  and Ca$^+$  are accelerated towards each other to form  a KCa$^+$ quasi-molecule with an oriented interatomic axis. 
In the presence of a linearly polarized laser, the behavior of the charge-exchange reaction rate as a function
of the angle  between the collision axis and the laser polarization direction can then reveal the presence of LICIs.
Prototype experiments  studying oriented collisions between two ultracold neutral atom clouds can be found in Ref.~\cite{Thomas2016}.

Our model of the former realization allows us to investigate the effect of the multiple ``pathways'' around the two LICIs that  occur in our system. These pathways lead to  interference patterns in the charge-exchange rate coefficient as function of the laser frequency and intensity. Moreover, shape resonances in the rate coefficients are observed as function  of the atom-ion collision energy. In a theoretical model, we can ``switch'' on and off the CI to further elucidate effect of a CI on the reaction dynamics.

It is important to note that there are many examples in the literature devoted to  increasing atom-ion charge-exchange reaction rates with external radiation \cite{Vitlina1975, Tang1976, Tang1977, George1982, Laughlin1985, Hsu1985, Ivanov2006, Hall2011, Petrov2017, Mills2019, Li2019, Li2020}. In most of these examples, the research  focused on colliding neutral and ionic atoms  where one or both of the colliding partners are prepared in optically excited electronic states,  which led to a rich variety of exit channels.   In  these studies,  the topology of a LICI was not invoked to explain the data or LICIs were not even present. For example, in a recent  paper in collaboration with K. Brown's group \cite{Li2020}, we demonstrated that  colliding neutral $^{39}$K atoms and optically excited $^{40}$Ca$^+$ ions confined in  spatially overlapping  magneto-optical  and Paul traps, respectively, exhibit  large charge-exchange rates. In this case LICIs were not present. The measured charge-exchange rate coefficients were in good agreement with our theoretical estimates. 

\section{Theoretical model}

Ultracold collisions of K and Ca$^+$, both in their doublet $^2$S electronic ground state, form molecular states that are superpositions of the excited singlet A$^1\Sigma^+$  and energetically lowest triplet a$^3\Sigma^+$ states \cite{Li2020}, where only the A$^1\Sigma^+$ component leads to radiative decay into bound states of the singlet ground X$^1\Sigma^+$ K$^+$Ca potential or the K$^+$($^1$S) and Ca($^1$S) continuum. This decay into the continuum corresponds to charge exchange with a small charge-exchange rate coefficient on the order of $10^{-14}$ cm$^3$/s \cite{Li2020}.

In the presence of laser light, the charge-exchange rate coefficient can be enhanced.
Figure \ref{AdiabaticPots} schematically shows our  theoretical model of the dressed K and Ca$^+$ collision.
 The X$^1\Sigma^+$ and A$^1\Sigma^+$ states are dressed with photons of energy $h\nu$ of the laser allowing for additional charge-exchange processes. Here, $\nu$ is the photon frequency and $h$ is the Planck constant. In Fig.~\ref{AdiabaticPots}(a)  the potential of the A$^1\Sigma^+$ state dressed with one photon crosses the potential of the $\tilde {\rm X}^1\Sigma^+$ state  twice when the photon wavelength is 716 nm. For a photon wavelength of  720 nm in Fig.~\ref{AdiabaticPots}(b), we create a glancing or Renner-Teller intersection \cite{Yarkony1998} of A$^1\Sigma^+$ and $\tilde {\rm X}^1\Sigma^+$ state potentials with a quadratic dependence  on the nuclear coordinates.  
The  colored surface plots in these panels show that the  two intersections and the one glancing  intersection, respectively,  are conical intersections in the $R\mhyphen\theta$ plane for the dressed adiabatic potentials that include the atom-light coupling.  In this article, the continuous-wave  laser light is linearly polarized along the space-fixed $z$ axis and
Fig.~\ref{AdiabaticPots}(c) shows the angle $\theta$ between the inter-atomic axis and the polarization of the laser light.
 
The dressed-state Hamiltonian for our system is
\begin{equation}
H =-\frac{\hbar^2}{2\mu_r}\frac{d^2}{dR^2} + \frac{{\bf L}^2}{2\mu_rR^2} + 
V^{\rm mol}(R) + V^{\rm rad}({\bf R}) +H_{\rm laser},
\label{eq:ham}
\end{equation}
where $R$ is the internuclear separation, $\mu_r$ is the reduced mass, ${\bf L}$ is the molecular orbital angular momentum operator with eigenstates $|\ell m_\ell\rangle$ and projection quantum number $m_\ell$ defined with respect to our space-fixed $z$ axis, $V^{\rm mol}(R)$ is the electronic Hamiltonian 
\begin{eqnarray}
   V^{\rm mol}(R)&=& \left ( 
        \begin{array}{cc}
            V_1(R)  &  0\\
            0 & V_2(R)
        \end{array}
        \right) \,,
        \end{eqnarray}
where $V_1(R)$ and $V_2(R)$ are the isotropic potential energies of the singlet X$^1\Sigma^+$ and A$^1\Sigma^+$ electronic states, respectively. Their electronic wavefunctions will be denoted by $|1\rangle$ and $|2\rangle$.
We use the potentials regarding their  computations and data tables are given in Supplemental Material. Note that
for $R\to\infty$, we have $V_2(R)-V_1(R)\to \Delta\equiv +hc\times 14296.114$ cm$^{-1}$ derived from neutral K and neutral Ca ionization energies found in Ref.~\cite{nist}.  The dissociation energy or depth and equilibrium separation of the A$^1\Sigma^+$ state potential are $D_{\rm e}=hc\times 1090$ cm$^{-1}$ and $R_{\rm e}=12.9a_0$. Here, $a_0$ is the Bohr radius. The depth of the X$^1\Sigma^+$  potential is just over four times larger than that of the A$^1\Sigma^+$ state.

The fourth term in Eq.~(\ref{eq:ham}), $V^{\rm rad}({\bf R})$, describes the electric dipole interaction with the laser field. 
In a body-fixed coordinate system, it is proportional to $\cos\theta$,
where $\theta\in[0,\pi]$ is the angle between the internuclear axis and the polarization of the laser field.
Hence, the molecule-field interaction  is anisotropic. In the rotating wave approximation,
\begin{equation}
    V^{\rm rad}({\bf R}) = -d(R) \sqrt{2 \pi h\nu/V} \cos\theta  \left[ a^\dagger |1\rangle \langle 2| + h.c.\right]\,,
    \label{eq:rad}
\end{equation}
where $d(R) $ is the $R$-dependent molecular electronic transition dipole moment between the X$^1\Sigma^+$ and A$^1\Sigma^+$  states and abbreviation $h.c.$ stands for the hermite conjugate.
Finally, $H_{\rm laser}=h\nu a^\dagger a$ describes the energy in our laser field with photon creation and annihilation operators $a$ and $a^\dagger$ in volume $V$, respectively. Its eigenstates  are $|n\rangle$ with energy $nh\nu$ for non-negative integers or photon numbers $n$. 
The operator $V^{\rm rad}({\bf R})$  has non-zero matrix elements between states that differ
by one photon number.  Effects of the permanent dipole moments of the  X$^1\Sigma^+$ and A$^1\Sigma^+$ states are neglected. They do not lead to charge exchange. We use the transition dipole moment from Ref.~\cite{Li2020}.
See Supplemental Material for a description and a data table.
Potential energy curves, spectroscopic constants, and transition electric dipole moments of K$^+$Ca were also determined and analyzed  in Ref.~\cite{Tomza2020}.

For  coupled-channels calculations, we use the basis $|i;\ell m_\ell ; n\rangle\equiv|i\rangle |\ell m_\ell\rangle  |n\rangle$ with $i=1,2$.
In this basis ${\bf L}^2$, $V_{\rm mol}$, and $H_{\rm laser}$ are diagonal. Only, $V^{\rm rad}({\bf R})$ couples basis functions with
  matrix elements 
\begin{eqnarray}
\lefteqn{ \langle 1;\ell m_\ell;n+1|V^{\rm rad}({\bf R})|2; \ell'm'_\ell;n\rangle} \label{eq:dressed} \\
   &=& -d(R)\sqrt{\frac{2 \pi I} {c}} \sqrt{\frac{2\ell'+1}{2\ell+1}}C^{\ell m_\ell}_{10,\ell'm'_\ell}C^{\ell0}_{10,\ell'0}\,,
              \nonumber
\end{eqnarray}
where $I$ is the laser intensity and $C^{jm}_{j_1m_1,j_2m_2}$ are Clebsch-Gordan coefficients. For our laser polarization $m'_\ell=m_\ell$
and the matrix element is only non-zero when $\ell+\ell'$ is odd. The dressed state picture, pioneered by C. Cohen-Tannoudji {\it et al.} \cite{CCT} and as used here, is for large photon number $n\ggg 1$  equivalent to Floquet theory based on a classical time-dependent description of light. For $n\ggg 1$, it is appropriate to replace the photon number dependence of the matrix element in Eq.~(\ref{eq:dressed}) by the mean photon number (per unit area) in the laser  and thus by  laser intensity $I$ as it is proportional to the mean photon number. As we will show the change in photon number due to the molecular processes is small even at our largest laser intensities.

The charge-exchange rate coefficient from $|1\rangle$ with $n$ photons and ultracold collision energy $E$ is given by
\begin{equation}
K = \frac{\hbar \pi}{\mu_r k} \sum_{\ell,\ell'=0}^{\ell_{\rm max}}\sum_{m_{\ell}}\sum_{n'=n-\delta n}^{n+\delta n}  \left| T_{2,\ell' m_{\ell},n'\leftarrow1,\ell,m_{\ell}, n} \right|^2,
\label{sum}
\end{equation}
where $k=\sqrt{2\mu_r E/\hbar^2}$ is collisional wave vector, $m_{\ell}$ runs from $-{\rm min}(\ell, \ell')$ to
${\rm min}(\ell, \ell')$, and $\hbar$ is the reduced Planck constant.
The quantities $T_{f\leftarrow i}$  are  T-matrix elements obtained from  scattering solutions
of Eq.~(\ref{eq:ham}), where the photon numbers changes by no more that  $\delta n$=1 or 2, 
and  partial waves $\ell$ from $0$ to ${\ell_{\rm max}=6}$ are coupled, sufficient for convergence at collision energies $E$ up to $k\times1$ mK and laser intensities up to $10^{8}$ W/cm$^2$.
Here, $k$ is the Boltzmann constant.
We find that the main contribution to $K$ comes from
$T$-matrix elements with $n'=n+1$, corresponding to the transition
 $|1\rangle +  nh\nu \rightarrow |2\rangle +  (n+1)h\nu$. 

In this paper, we focus on  photon energies between $hc\times 13600$ cm$^{-1}$ and $\Delta=hc\times14296.114$ cm$^{-1}$, to ensure the presence of up to two LICIs between the X$^1\Sigma^+$ and A$^1\Sigma^+$ potentials, as  shown  Fig.~\ref{AdiabaticPots}. 
The LICIs are best visible in the surface graphs of Figs.~\ref{AdiabaticPots}(a) and (b) corresponding to the eigenvalues of operator $V^{\rm mol}(R)+V^{\rm rad}({\bf R})$ as functions of $R$ and $\theta$. The LICIs occur at $\theta=\pi/2$ with
 separations $R_1$ and $R_2$, where $R_1\le R_2$. For the Renner-Teller intersection $R_1=R_2> R_{\rm e}$.

\section{Results and Discussions}

\begin{figure} 
\includegraphics[scale=0.30,trim=10 10 0 20,clip]{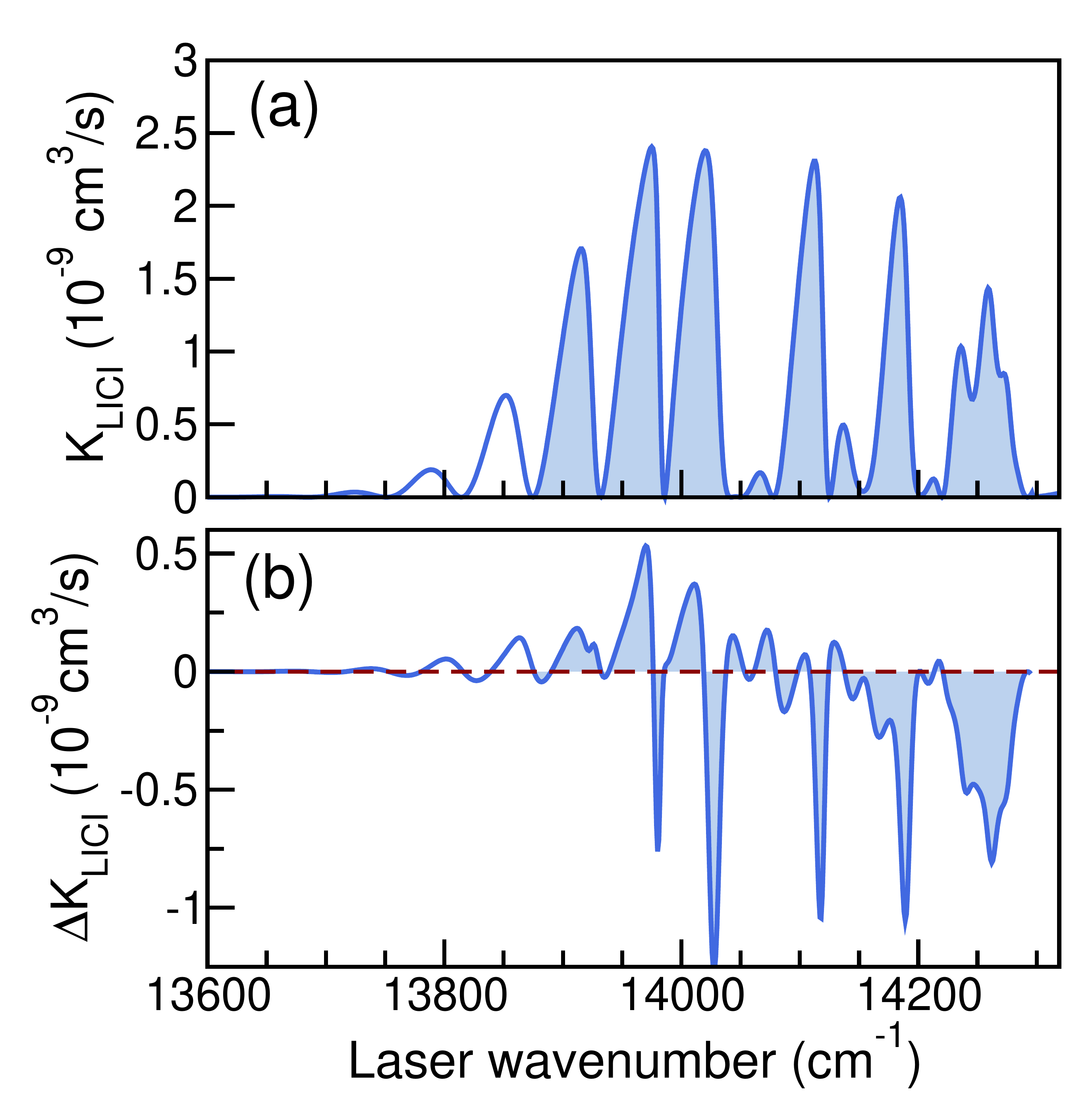}
\caption{a) The charge-exchange  rate coefficient K$_{\rm LICI}$  as a function of laser wavenumber at a collision energy of $E/k=7$ $\mu$K and a laser intensity of $I=10^8$ W/cm$^2$. The shaded light-blue region indicates 
the frequency region where  interferences between pathways around the two conical intersections occur. b) The difference $\Delta K_{\rm LICI}=K_{\rm no\mhyphen LICI}-K_{\rm LICI}$ as a function of laser wavenumber 
at $E/k=7$ $\mu$K and $I=10^8$ W/cm$^2$. Here,  $K_{\rm no\mhyphen LICI}$ and $K_{\rm LICI}$ are charge-exchange rate coefficients computed without and with the conical intersections, respectively.   }
\label{FreqDep}
\end{figure}

{\bf Interference patterns between pathways around two LICIs}. We begin by investigating the effect of LICIs on rate coefficients of the charge-exchange reaction $K_{\rm LICI}$ by calculating the rate coefficient around a $hc\times 800$ cm$^{-1}$ range of photon energies below  the difference in dissociation energy of the X$^1\Sigma^+$ and A$^1\Sigma^+$ states. A representative example is shown in Fig.~\ref{FreqDep}(a) for a collision energy of $E/k=7$ $\mu$K and a laser intensity of $10^8$ W/cm$^2$. 
The laser wavenumber is increased from $\nu/c=13600$ cm$^{-1}$ to 14296.114 cm$^{-1}=\Delta/hc$, where our system goes from having no LICI  to one where two LICIs exist between the A$^1\Sigma^+$ and $\tilde{\rm X}^1\Sigma^+$ potentials.  
We observe  regular and irregular St\"uckelberg oscillations, where the rate coefficient alternates between near zero values and maxima as function of $\nu$. 
Below $\nu/c=13600$ cm$^{-1}$ the light-induced charge-exchange  rate  coefficient is below $\sim10^{-14}$ cm$^3$/s and  radiative decay  studied in Ref.~\cite{Li2020} becomes the dominant means for charge exchange. For $h\nu>
\Delta$   light-induced rate coefficient $K_{\rm LICI}$ is negligibly small as transition
 $|1\rangle +  nh\nu \rightarrow |2\rangle +  (n+1)h\nu$ is energetically forbidden.
 
For photon energies between $hc\times 13600$ cm$^{-1}$ and $hc\times13896$ cm$^{-1}$ we observe St\"uckelberg oscillations with increasing amplitude corresponding to cases where we only have avoided crossings between A$^1\Sigma^+$ and $\tilde{\rm X}^1\Sigma^+$ potentials. The amplitude of the oscillations increases because the avoided crossing between the potentials become narrower.
Starting from laser energies  $hc\times 13896$ cm$^{-1}$, shaded in Fig.~\ref{FreqDep}(a), two conical intersections are present and the oscillation pattern of  $K_{\rm LICI}$ is significantly distorted or irregular due to interferences between multiple pathways around the two conical intersections. 

To support our claim that  the irregular St\"uckelberg oscillations are due to  LICIs, we first analyze the charge-exchange rate coefficient for selected laser frequencies in this frequency window.
Specifically, we analyze  perturbative  transition matrix elements between scattering wavefunctions in channels
$|1;\ell m_\ell;n\rangle$ and $|2;\ell' m'_\ell;n+1\rangle$ as function of laser frequency
and locate the origin of the complex constructive and destructive interferences corresponding to large and small charge-exchange rate coefficients. 
We  define the {\it partial} matrix element or integral
\begin{equation}
     M(R) = \int_0^R {\rm d}r \psi_{\rm X}(r) d(r) \phi_{\rm A}(r)\,,
\end{equation}
where real-valued $\phi_{\rm A}(r)$ is the single-channel $\ell=0$ radial scattering wave function for the 
A$^1\Sigma^+$ potential at initial collision energy $E_{\rm initial}$ and real-valued
$\psi_{\rm X}(r)$ is the single-channel $\ell=1$ radial scattering wave function for the $\tilde{\rm X}^1\Sigma^+$ potential at final collision energy $E_{\rm final}=\Delta-h\nu+E_{\rm initial}$.
For weak laser intensities, the charge-exchange rate coefficient $K_{\rm LICI}$ is proportional to the square of $M(R)$ for $R\to\infty$. Constructive or destructive interference implies large or small $|M(R\to\infty)|$, respectively.

\begin{figure} 
\includegraphics[scale=0.25,trim=0 0 0 0,clip]{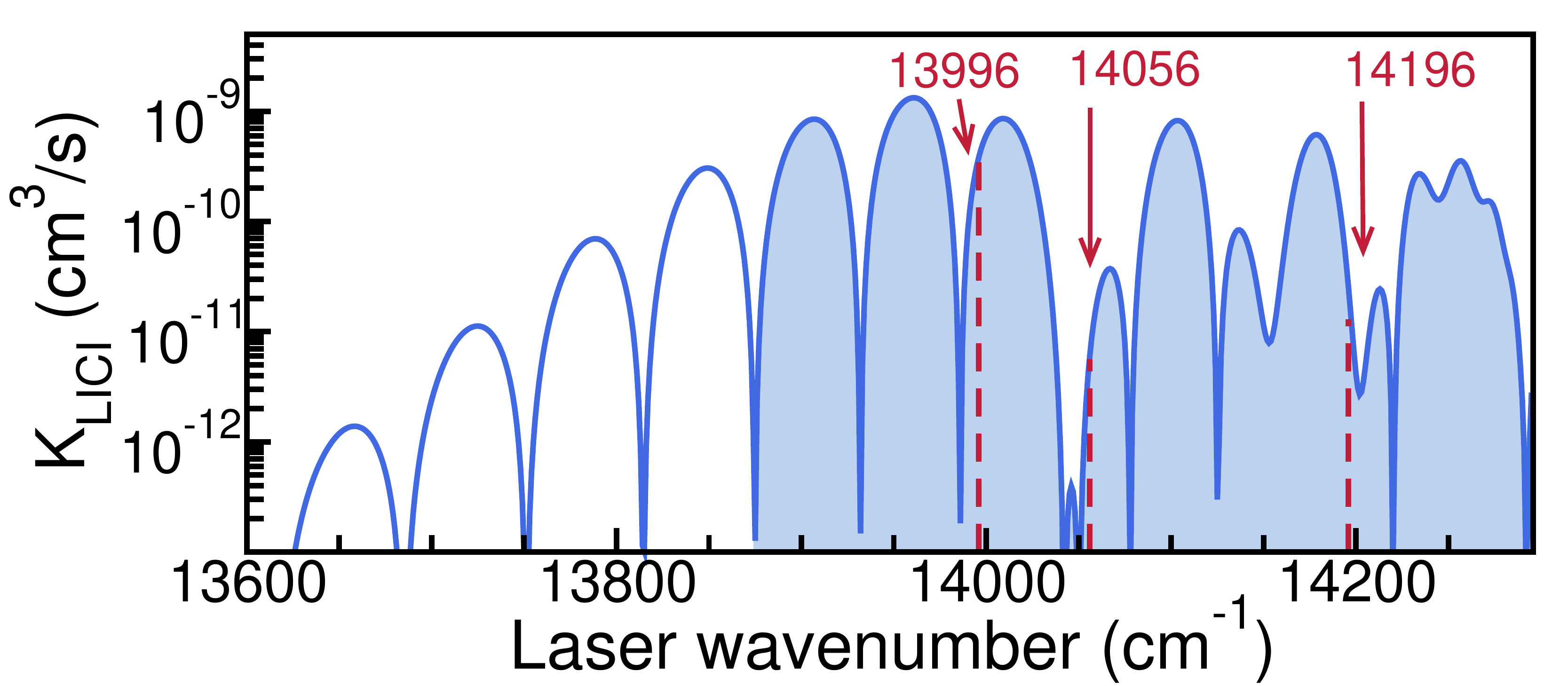}
\caption{The charge-exchange  rate coefficient K$_{\rm LICI}$  as a function of laser wavenumber at $E/k=1$ $\mu$K and  $I=10^8$ W/cm$^2$. The shaded light-blue region again indicates 
the frequency region where  interferences between pathways around the two conical intersections occur.  Wavenumbers 13996 cm$^{-1}$, 14056 cm$^{-1}$, and 14196 cm$^{-1}$ with dashed red lines mark  dressing conditions leading to either  constructive or near destructive interference.}
\label{fig:Ekinis1}
\end{figure}

\begin{figure} 
\includegraphics[scale=0.32,trim=10 10 0 0,clip]{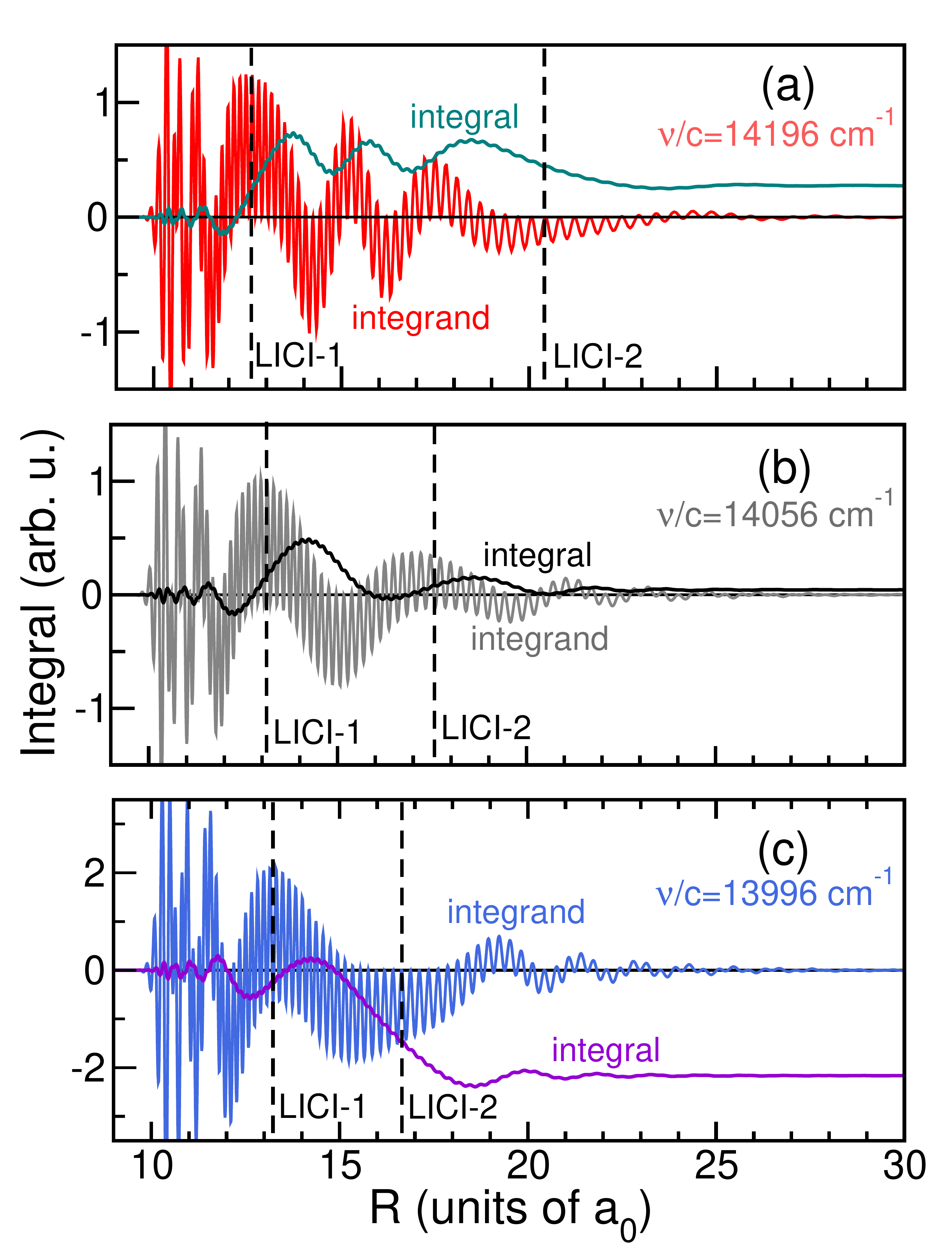}
\caption{Explanation of the irregular St\"uckelberg oscillations in the K+Ca$^+$ charge-exchange rate coefficient at a collision energy of $k\times 1$ $\mu$K shown in Fig.~\ref{fig:Ekinis1} when the dressed-state potentials have two LICIs. Panels (a), (b), and (c) show the  integrand and
partial integral  of  vibrationally averaged electronic transition dipole moments as functions of atom-atom separation $R$ for laser wavenumbers $\nu/c= 14196$ cm$^{-1}$, 14056 cm$^{-1}$, and  13996 cm$^{-1}$, respectively. 
The vertical black dashed lines labeled LICI-1 and LICI-2 in each panel correspond to the KCa$^+$ separations of the LICIs. 
The scale on the $y$ axes of the panels are different and should not be compared. }
\label{Path}
\end{figure}

We begin the analysis by showing rate coefficient $K_{\rm LICI}$ at  collision energy of $E/k=1$ $\mu$K
in Fig.~\ref{fig:Ekinis1} on a logarithmic scale. The interference pattern is similar to that observed in Fig.~\ref{FreqDep}
at a larger collision energy. Figure~\ref{Path} then shows the partial matrix element $M(R)$ and the integrand $\psi_{\rm X}(R) d(R) \phi_{\rm A}(R)$   as functions of $R$  for an initial collision energy of $E_{\rm initial}=1$ $\mu$K  when the wavenumbers of the dressing laser are $\nu/c= 14196$ cm$^{-1}$, 14056 cm$^{-1}$, and 13996 cm$^{-1}$, respectively. 
The product $d(R) \phi_{\rm A}(R)$ is the same for the three cases and only $\psi_{\rm X}(R) $ changes.

For the largest photon energy, shown in panel (a), the system has two  well-separated conical intersections located at separations $R_1$ and $R_2$ with $R_1< R_2$. They are labelled LICI-1 and LICI-2 in the figure.
We also observe that the integrand is a rapidly oscillating function with $R$ and  that it approaches zero for large $R$ as $d(R)\to 0$ for $R\to\infty$.
Hence, we expect that cancellations but also non-zero averages will occur in the calculation of $M(R)$.
Near  $R=R_1$ and $R_2$, where the local kinetic energies for $\tilde{\rm X}^1\Sigma^+$ and A$^1\Sigma^+$ states are close to equal, however, the oscillations in the integrand occur around a non-zero average
and, indeed, for $R$ near $R_1$ the integral is seen to rapidly increase. Near $R=R_2$ the integral decreases somewhat.
For all other $R$, $M(R)$ oscillates around a stable value.
Thus, we realize that for  $\nu/c= 14196$ cm$^{-1}$ and $R\to\infty$ the contributions to $M(R)$ from the two LICIs partially cancel. 
 
For the data at $\nu/c=14056$ cm$^{-1}$  in Fig.~\ref{Path}(b)  the contributions from the two LICIs nearly cancel each other.
There is destructive interference between the LICIs.
For the data in panel (c) and $\nu/c=13996$ cm$^{-1}$, the radial separations of the two CIs  are closest of all three cases and the contributions
from the two LICIs are harder to separate. Still,  the contribution to $M(R)$ near LICI-1 averages to near zero,
while the contribution from LICI-2 is  large. From similar figures at other kinetic energies and laser frequencies, not shown, we can understand 
the patterns seen in Fig.~\ref{FreqDep}(a) as function of laser frequency.

We have also performed a second type of analysis to confirm that the light-induced conical intersections influence the charge-exchange process. We ``removed'' the conical intersections by replacing the anisotropic $\theta$ dependence in Eq.~(\ref{eq:rad}) with an isotropic one. In fact, we made the substitution 
\begin{equation}
   \cos\theta \to \frac{1}{\sqrt{3}}
\end{equation}
in Eq.~(\ref{eq:rad}). The molecule-field interaction is now isotropic and only couples channels with the same
partial wave $\ell$ and $m_{\ell}$ quantum numbers.

Figure~\ref{FreqDep}(b) shows the difference of the charge-exchange rate coefficients  
$K_{\rm no\mhyphen LICI}$ and $K_{\rm LICI}$ 
as a function of laser frequency at $E/k=1$ $\mu$K and  $I=10^8$ W/cm$^2$.
Here, $K_{\rm no\mhyphen LICI}$ and $K_{\rm LICI}$ correspond to rate coefficients computed without and with conical intersections, respectively.
The difference in rate coefficients can be of the same order of magnitude as $K_{\rm LICI}$  and be as large as $\pm 10^{-9}$ cm$^3$/s in the frequency window where the conical interactions are present.
For ultracold atom and ion experiments, a rate coefficient of order of $10^{-9}$ cm$^3$/s is large and easily detectable.

\begin{figure} 
\includegraphics[scale=0.32,trim=0 10 0 0,clip]{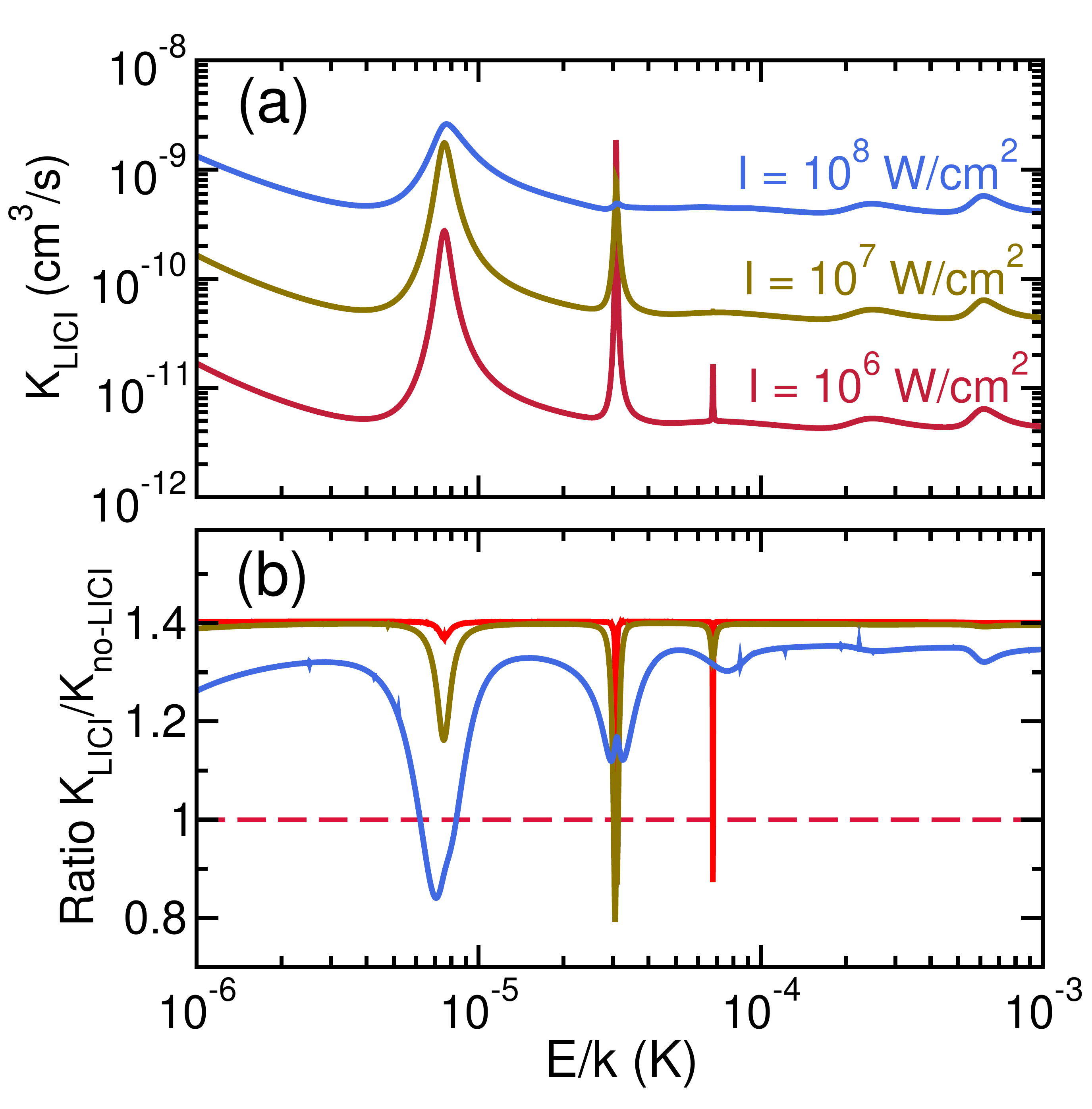}
\caption{The charge-exchange rate coefficient  K$_{\rm LICI}$ computed with  the CIs present (panel a) and the ratio K$_{\rm LICI}/K_{\rm no\mhyphen LICI}$ (panel b) as functions of atom-ion collision energy at laser wavenumber $\nu$/c=13960 cm$^{-1}$ and at  laser intensities of $I=10^6$ W/cm$^2$ (red curve),  $10^7$ W/cm$^2$ (gold), and $10^8$ W/cm$^2$ (blue).  }
\label{LICI_noLICI_E}
\end{figure} 

Finally, we examine  charge-exchange rate coefficients between K and Ca$^+$  as functions of collision energy $E$ at $\nu$/c=13960 cm$^{-1}$ and  laser intensities $10^6$ W/cm$^2$, $10^7$ W/cm$^2$, and $10^8$ W/cm$^2$.
We remain in the ultracold collision energy domain below $k\times 1$ mK.
This case corresponds to a situation where $K_{\rm LICI}$ is close to maximal when $E\to0$ and the pair of interfering conical intersections are located close together. Figure~\ref{LICI_noLICI_E}(a) shows  rate coefficients $K_{\rm LICI}$ where the LICIs present. We observe that both wide and narrow resonance features are present. 
Further analysis has shown that these resonances are due to shape resonances behind  a $\ell$-wave centrifugal barrier of the long-range $-C_4/R^4+\hbar^2\ell(\ell+1)/(2\mu_r R^2)$ potential between a neutral atom and an ionic atom. 
Here, $C_4$ is the polarization coefficient proportional to the static polarizability of the K atom.
These resonances significantly enhance the charge-exchange process. A broader resonance has an energy closer to the top of the corresponding centrifugal barrier. 
The rate coefficients increase approximately linearly with laser intensity, although for the largest intensity of $10^8$ W/cm$^2$
deviations are visible. For example, the resonances have noticeably broadened, which can be used as a signal that CIs are present.

Figure~\ref{LICI_noLICI_E}(b) shows the ratio $K_{\rm LICI}/K_{\rm no\mhyphen LICI}$ of  $\nu/c$=13960 cm$^{-1}$ rate coefficients $K_{\rm LICI}$ and $K_{\rm no\mhyphen LICI}$ calculated  when the LICIs are and are not present, respectively.   
We observe that for the smaller laser intensities the ratio is about 1.4 independent of collision energy and laser intensity, except near the shape resonances.
At the higher laser intensity of $10^8$ W/cm$^2$ the background value of the ratio of rate coefficients also changes in a measurable way, although the width and strength of the resonances are again affected more dramatically. 

{\bf In Summary}: We have computed charge-exchange rate coefficients of colliding K atoms and Ca$^+$ ions confined in hybrid atom-ion traps at  ultracold temperatures  in the presence of a near resonant laser beam that initiates non-adiabatic coupling between the ground and first excited electronic potential energy surfaces of this ionic system. The underlying time-independent coupled-channels calculations  allowed us to treat the non-adiabatic nuclear dynamics, dominated by two light-induced conical intersections, exactly.  In particular, our  state-of-the-art calculations showed the presence of interferences between pathways around two conical intersections. 
We have shown that these laser-induced conical intersections can significantly increase the atom-ion chemical reactivity at 
ultracold collision energies and relatively small intensities of the dressing laser. The expected rate coefficients should be easily detectable in current experimental setups.
We also investigated  charge exchange when the conical intersections were removed. The results of those  studies  showed that the conical intersections lead to noticeable differences in the reactive charge-exchange reaction.

\vspace*{0.5cm}
\noindent
{\bf \large{Acknowledgements}}\\
\noindent
Work at Temple University is supported by the U.S. Air Force Office of Scientific Research Grants No. FA9550-21-1-0153 and
No. FA9550-19-1-0272 and the NSF Grant No. PHY-1908634.

\bibliographystyle{achemso}
\bibliography{BibTexLibraryKotochigova_mod}

\end{document}


\title{Supplemental Material: Quantum Control of Atom-Ion Charge Exchange via Light-induced Conical Intersections}

\author{Hui Li}
\thanks{Present address: JILA, University of Colorado, Boulder, Colorado 80309, USA}
\address{Department of Physics, Temple University, Philadelphia, Pennsylvania 19122, USA}
\author{Ming Li}
\address{Department of Physics, Temple University, Philadelphia, Pennsylvania 19122, USA}
\author{Alexander Petrov}
\address{Department of Physics, Temple University, Philadelphia, Pennsylvania 19122, USA}
\author{Eite Tiesinga}
\address{Joint Quantum Institute, National Institute of Standards and Technology and University of Maryland, Gaithersburg, Maryland
20899, USA}
\author{Svetlana Kotochigova}
\email{skotoch@temple.edu}
\address{Department of Physics, Temple University, Philadelphia, Pennsylvania 19122, USA}

\maketitle

\subsection{The electronic structure of the KCa$^+$ molecule}

In this paper we focus on two electronic potentials of the KCa$^+$ molecule, the ground state X$^1\Sigma^+$  potential and the excited A$^1\Sigma^+$ potential. We calculated these potentials using the non-relativistic multi-reference configuration 
interaction (MRCI) method as implemented in the MOLPRO program \cite{Werner2012}. These calculations provide us with the electronic transition dipole moment between the two states as well.
The inner-shell electrons for both K and Ca are described by the Stevens-Basch-Krauss-Jansien-Cundari (SBKJC)
effective core potentials (ECP)~\cite{Stevens1992}, leaving only two valence electrons in the active space. The 
polarization of the effective cores is modeled via the core polarization potential (CPP) described in
Ref.~\cite{Muller1984}. For the CPP, we use the static dipole polarizabilities of K$^+$ and Ca$^{2+}$, which are $5.472$
a.u.~\cite{Wilson1970} and $3.522$ a.u.~\cite{Coker1976} respectively. Here, a.u. is an abbreviation  for atomic unit. 
M\"uller/Meyer type cutoff functions are used with their exponents fitted to the asymptotic energies at large 
 separation $R$, which yields the values $0.383$ a.u. and $0.465$ a.u. for K and Ca, respectively. For K we use the large 
7s5p7d2f uncontracted Gaussian basis set from Ref.~\cite{Aymar2005}. For Ca we use the contracted and 
optimized 9s7p6d/[7s6p6d] basis set from Ref.~\cite{Czuchaj2000} augmented by three sets of f                  
and one set of g polarization functions~\cite{Czuchaj2003}. The contractions are specified in the square bracket. 
The multi-configurational self-consistent field (MCSCF) \cite{Werner1985a,Werner1985b} method is  used  
to obtain molecular orbitals, followed by a full configuration interaction calculation (MRCI) with two-electron excitations.

\begin{figure}
 \includegraphics[width=\columnwidth,trim=10 10 0 0,clip]{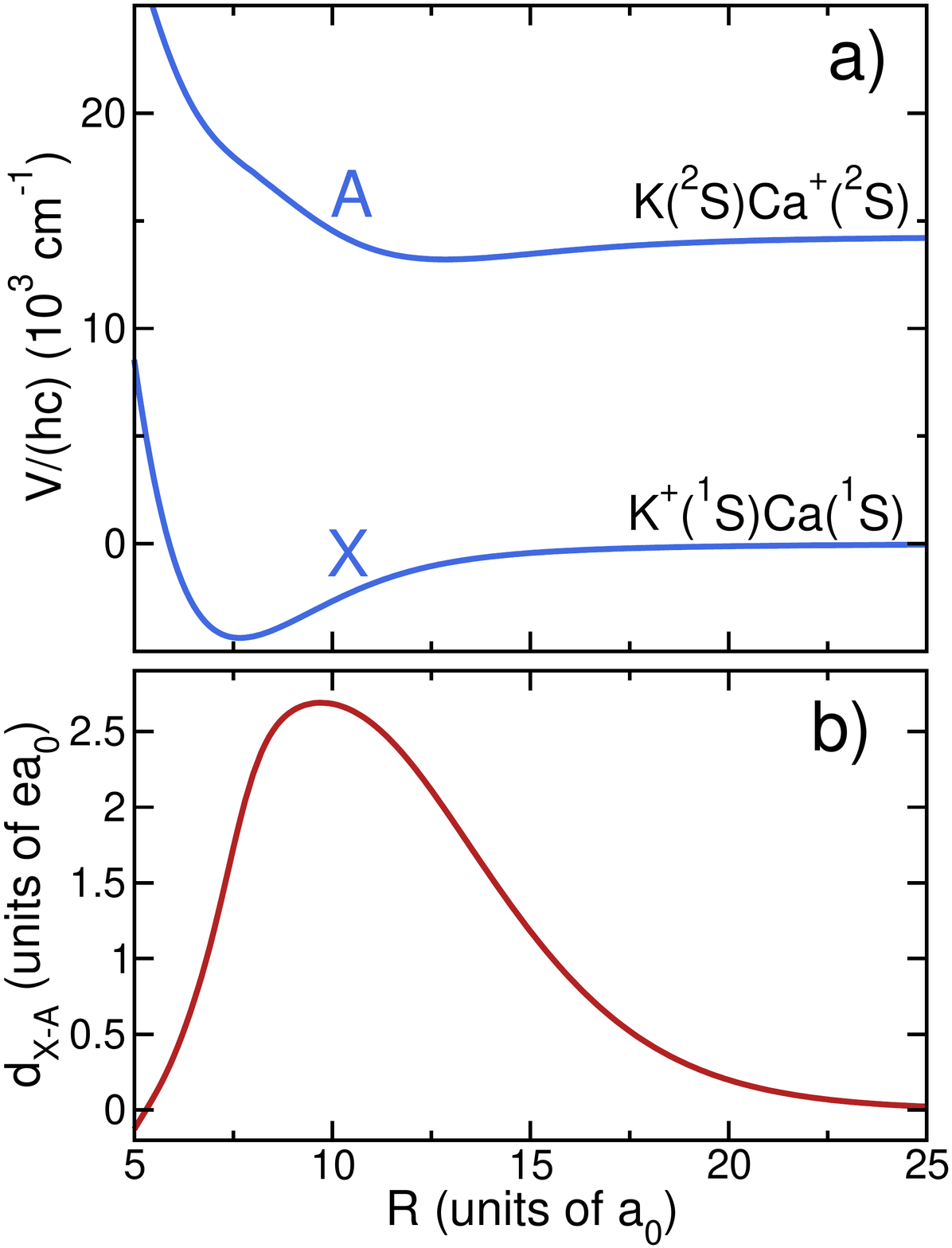}
 \caption{(a) Non-relativistic X$^1\Sigma^+$ and A$^1\Sigma^+$  Born-Oppenheimer potentials of KCa$^+$ as functions of interatomic separation $R$.  
  Asymptotic atomic configurations are given on the right hand side of the panel. The zero of energy is at the ${\rm K}^+(^1{\rm S})+{\rm Ca}(4{\rm s}^2\, ^1{\rm S})$
  dissociation limit. (b) $R$-dependent transition dipole moment between the X$^1\Sigma^+$ and A$^1\Sigma^+$ states.}
\label{pes_tdm}
\end{figure}

Figure \ref{pes_tdm}(a) shows the  MRCI potential energies for the X$^1\Sigma^+$  and A$^1\Sigma^+$ states as a function of interatomic separation.  Figure \ref{pes_tdm}(b) shows the electric transition dipole moment between these  states. 
Data for this figure are found in Tables \ref{tab:one},  \ref{tab:two}, \ref{tab:three}, and \ref{tab:four}.

\begin{table*} 
\caption{Part I: The  X$^1\Sigma^+$ and A$^1\Sigma^+$  potential energy curves of KCa$^+$.}
\label{tab:one}
\begin{tabular}{rrr|rrr|rrr|rrr}
\hline
	$R/a_0$  &   $V_{\rm X}/hc$  &   $V_{\rm A}/hc$  &
	$R/a_0$  &   $V_{\rm X}/hc$  &   $V_{\rm A}/hc$  &
	$R/a_0$  &   $V_{\rm X}/hc$  &   $V_{\rm A}/hc$  &
	$R/a_0$  &   $V_{\rm X}/hc$  &   $V_{\rm A}/hc$  \\
	& (cm$^{-1}$) & (cm$^{-1}$) & 
	& (cm$^{-1}$) & (cm$^{-1}$) & 
	& (cm$^{-1}$) & (cm$^{-1}$) & 
	& (cm$^{-1}$) & (cm$^{-1}$)  
        \\
\hline
 4.0 &  21446.976 &  35486.971 & 	11.0 &  $-$1863.422 &  13696.072 & 	17.6 &   $-$209.574 &  13851.081 & 	24.2 &    $-$54.460 &  14191.014 \\ 
 4.1 &  20488.572 &  34659.956 & 	11.1 &  $-$1793.755 &  13636.918 & 	17.7 &   $-$204.428 &  13862.210 & 	24.3 &    $-$53.539 &  14192.888 \\ 
 4.2 &  19379.026 &  33838.999 & 	11.2 &  $-$1726.303 &  13582.236 & 	17.8 &   $-$199.447 &  13873.048 & 	24.4 &    $-$52.637 &  14194.719 \\ 
 4.3 &  18146.077 &  33032.147 & 	11.3 &  $-$1661.070 &  13531.919 & 	17.9 &   $-$194.623 &  13883.598 & 	24.5 &    $-$51.755 &  14196.508 \\ 
 4.4 &  16820.757 &  32243.531 & 	11.4 &  $-$1598.049 &  13485.849 & 	18.0 &   $-$189.951 &  13893.866 & 	24.6 &    $-$50.891 &  14198.256 \\ 
 4.5 &  15434.699 &  31474.418 & 	11.5 &  $-$1537.225 &  13443.898 & 	18.1 &   $-$185.423 &  13903.856 & 	24.7 &    $-$50.046 &  14199.965 \\ 
 4.6 &  14017.881 &  30724.355 & 	11.6 &  $-$1478.577 &  13405.928 & 	18.2 &   $-$181.036 &  13913.573 & 	24.8 &    $-$49.218 &  14201.635 \\ 
 4.7 &  12597.047 &  29992.166 & 	11.7 &  $-$1422.076 &  13371.799 & 	18.3 &   $-$176.782 &  13923.024 & 	24.9 &    $-$48.407 &  14203.268 \\ 
 4.8 &  11194.851 &  29276.652 & 	11.8 &  $-$1367.690 &  13341.361 & 	18.4 &   $-$172.658 &  13932.212 & 	25.0 &    $-$47.613 &  14204.865 \\ 
 4.9 &   9829.609 &  28577.007 & 	11.9 &  $-$1315.378 &  13314.465 & 	18.5 &   $-$168.658 &  13941.145 & 	25.1 &    $-$46.835 &  14206.427 \\ 
 5.0 &   8515.474 &  27892.992 & 	12.0 &  $-$1265.098 &  13290.954 & 	18.6 &   $-$164.777 &  13949.828 & 	25.2 &    $-$46.073 &  14207.954 \\ 
 5.1 &   7262.860 &  27224.956 & 	12.1 &  $-$1216.801 &  13270.673 & 	18.7 &   $-$161.012 &  13958.266 & 	25.3 &    $-$45.327 &  14209.448 \\ 
 5.2 &   6078.972 &  26573.759 & 	12.2 &  $-$1170.438 &  13253.463 & 	18.8 &   $-$157.357 &  13966.466 & 	25.4 &    $-$44.596 &  14210.909 \\ 
 5.3 &   4968.351 &  25940.656 & 	12.3 &  $-$1125.955 &  13239.167 & 	18.9 &   $-$153.809 &  13974.434 & 	25.5 &    $-$43.879 &  14212.339 \\ 
 5.4 &   3933.382 &  25327.141 & 	12.4 &  $-$1083.297 &  13227.629 & 	19.0 &   $-$150.364 &  13982.176 & 	25.6 &    $-$43.177 &  14213.738 \\ 
 5.5 &   2974.744 &  24734.819 & 	12.5 &  $-$1042.407 &  13218.694 & 	19.1 &   $-$147.018 &  13989.697 & 	25.7 &    $-$42.489 &  14215.107 \\ 
 5.6 &   2091.785 &  24165.268 & 	12.6 &  $-$1003.226 &  13212.208 & 	19.2 &   $-$143.768 &  13997.004 & 	25.8 &    $-$41.815 &  14216.446 \\ 
 5.7 &   1282.834 &  23619.935 & 	12.7 &   $-$965.697 &  13208.022 & 	19.3 &   $-$140.609 &  14004.103 & 	25.9 &    $-$41.154 &  14217.758 \\ 
 5.8 &    545.460 &  23100.053 & 	12.8 &   $-$929.759 &  13205.987 & 	19.4 &   $-$137.540 &  14010.999 & 	26.0 &    $-$40.507 &  14219.042 \\ 
 5.9 &   $-$123.325 &  22606.577 & 	12.9 &   $-$895.353 &  13205.962 & 	19.5 &   $-$134.556 &  14017.697 & 	26.1 &    $-$39.872 &  14220.299 \\ 
 6.0 &   $-$726.889 &  22140.151 & 	13.0 &   $-$862.421 &  13207.805 & 	19.6 &   $-$131.656 &  14024.205 & 	26.2 &    $-$39.249 &  14221.529 \\ 
 6.1 &  $-$1268.846 &  21701.095 & 	13.1 &   $-$830.904 &  13211.383 & 	19.7 &   $-$128.835 &  14030.527 & 	26.3 &    $-$38.639 &  14222.734 \\ 
 6.2 &  $-$1752.937 &  21289.407 & 	13.2 &   $-$800.744 &  13216.563 & 	19.8 &   $-$126.091 &  14036.668 & 	26.4 &    $-$38.041 &  14223.914 \\ 
 6.3 &  $-$2182.947 &  20904.775 & 	13.3 &   $-$771.887 &  13223.221 & 	19.9 &   $-$123.422 &  14042.635 & 	26.5 &    $-$37.454 &  14225.070 \\ 
 6.4 &  $-$2562.629 &  20546.606 & 	13.4 &   $-$744.275 &  13231.235 & 	20.0 &   $-$120.825 &  14048.432 & 	26.6 &    $-$36.878 &  14226.203 \\ 
 6.5 &  $-$2895.660 &  20214.046 & 	13.5 &   $-$717.856 &  13240.489 & 	20.1 &   $-$118.298 &  14054.065 & 	26.7 &    $-$36.314 &  14227.312 \\ 
 6.6 &  $-$3185.592 &  19906.013 & 	13.6 &   $-$692.577 &  13250.870 & 	20.2 &   $-$115.838 &  14059.538 & 	26.8 &    $-$35.760 &  14228.398 \\ 
 6.7 &  $-$3435.836 &  19621.221 & 	13.7 &   $-$668.387 &  13262.274 & 	20.3 &   $-$113.444 &  14064.857 & 	26.9 &    $-$35.217 &  14229.463 \\ 
 6.8 &  $-$3649.637 &  19358.205 & 	13.8 &   $-$645.238 &  13274.598 & 	20.4 &   $-$111.112 &  14070.026 & 	27.0 &    $-$34.685 &  14230.507 \\ 
 6.9 &  $-$3830.063 &  19115.346 & 	13.9 &   $-$623.082 &  13287.747 & 	20.5 &   $-$108.841 &  14075.050 & 	27.1 &    $-$34.162 &  14231.529 \\ 
 7.0 &  $-$3980.003 &  18890.891 & 	14.0 &   $-$601.873 &  13301.629 & 	20.6 &   $-$106.630 &  14079.933 & 	27.2 &    $-$33.649 &  14232.532 \\ 
 7.1 &  $-$4102.162 &  18682.981 & 	14.1 &   $-$581.567 &  13316.158 & 	20.7 &   $-$104.475 &  14084.680 & 	27.3 &    $-$33.146 &  14233.514 \\ 
 7.2 &  $-$4199.062 &  18489.672 & 	14.2 &   $-$562.123 &  13331.252 & 	20.8 &   $-$102.376 &  14089.296 & 	27.4 &    $-$32.652 &  14234.477 \\ 
 7.3 &  $-$4273.048 &  18308.971 & 	14.3 &   $-$543.500 &  13346.835 & 	20.9 &   $-$100.330 &  14093.783 & 	27.5 &    $-$32.168 &  14235.421 \\ 
 7.4 &  $-$4326.294 &  18138.879 & 	14.4 &   $-$525.659 &  13362.833 & 	21.0 &    $-$98.336 &  14098.147 & 	27.6 &    $-$31.692 &  14236.347 \\ 
 7.5 &  $-$4360.809 &  17977.438 & 	14.5 &   $-$508.563 &  13379.180 & 	21.1 &    $-$96.392 &  14102.391 & 	27.7 &    $-$31.225 &  14237.255 \\ 
 7.6 &  $-$4378.445 &  17822.790 & 	14.6 &   $-$492.178 &  13395.811 & 	21.2 &    $-$94.497 &  14106.519 & 	27.8 &    $-$30.767 &  14238.145 \\ 
 7.7 &  $-$4380.911 &  17673.232 & 	14.7 &   $-$476.468 &  13412.669 & 	21.3 &    $-$92.649 &  14110.535 & 	27.9 &    $-$30.317 &  14239.019 \\ 
 7.8 &  $-$4369.776 &  17527.262 & 	14.8 &   $-$461.403 &  13429.697 & 	21.4 &    $-$90.847 &  14114.442 & 	28.0 &    $-$29.875 &  14239.875 \\ 
 8.0 &  $-$4312.363 &  17261.643 & 	14.9 &   $-$446.951 &  13446.845 & 	21.5 &    $-$89.090 &  14118.244 & 	28.1 &    $-$29.442 &  14240.715 \\ 
 8.2 &  $-$4216.410 &  16957.669 & 	15.0 &   $-$433.083 &  13464.066 & 	21.6 &    $-$87.375 &  14121.944 & 	28.2 &    $-$29.016 &  14241.539 \\ 
 8.3 &  $-$4156.726 &  16815.563 & 	15.1 &   $-$419.772 &  13481.315 & 	21.7 &    $-$85.702 &  14125.545 & 	28.3 &    $-$28.598 &  14242.348 \\ 
 8.4 &  $-$4090.524 &  16673.014 & 	15.2 &   $-$406.991 &  13498.552 & 	21.8 &    $-$84.070 &  14129.050 & 	28.4 &    $-$28.188 &  14243.141 \\ 
 \hline
\end{tabular}
\end{table*}
 \begin{table*} 
\caption{Part II: The  X$^1\Sigma^+$ and A$^1\Sigma^+$  potential energy curves of KCa$^+$.}
\label{tab:two}
\begin{tabular}{rrr|rrr|rrr|rrr}
\hline
	$R/a_0$  &   $V_{\rm X}/hc$  &   $V_{\rm A}/hc$  &
	$R/a_0$  &   $V_{\rm X}/hc$  &   $V_{\rm A}/hc$  &
	$R/a_0$  &   $V_{\rm X}/hc$  &   $V_{\rm A}/hc$  &
	$R/a_0$  &   $V_{\rm X}/hc$  &   $V_{\rm A}/hc$  \\
	& (cm$^{-1}$) & (cm$^{-1}$) & 
	& (cm$^{-1}$) & (cm$^{-1}$) & 
	& (cm$^{-1}$) & (cm$^{-1}$) & 
	& (cm$^{-1}$) & (cm$^{-1}$)  
        \\
\hline
 8.5 &  $-$4018.671 &  16530.043 & 	15.3 &   $-$394.715 &  13515.742 & 	21.9 &    $-$82.477 &  14132.463 & 	28.5 &    $-$27.784 &  14243.920 \\ 
 8.6 &  $-$3941.963 &  16386.798 & 	15.4 &   $-$382.921 &  13532.849 & 	22.0 &    $-$80.922 &  14135.786 & 	28.6 &    $-$27.388 &  14244.684 \\ 
 8.7 &  $-$3861.128 &  16243.526 & 	15.5 &   $-$371.586 &  13549.844 & 	22.1 &    $-$79.404 &  14139.023 & 	28.7 &    $-$27.000 &  14245.433 \\ 
 8.8 &  $-$3776.837 &  16100.545 & 	15.6 &   $-$360.687 &  13566.699 & 	22.2 &    $-$77.922 &  14142.175 & 	28.8 &    $-$26.618 &  14246.169 \\ 
 8.9 &  $-$3689.701 &  15958.224 & 	15.7 &   $-$350.206 &  13583.388 & 	22.3 &    $-$76.475 &  14145.245 & 	28.9 &    $-$26.242 &  14246.892 \\ 
 9.0 &  $-$3600.282 &  15816.961 & 	15.8 &   $-$340.122 &  13599.890 & 	22.4 &    $-$75.062 &  14148.237 & 	29.0 &    $-$25.874 &  14247.601 \\ 
 9.1 &  $-$3509.091 &  15677.167 & 	15.9 &   $-$330.417 &  13616.184 & 	22.5 &    $-$73.681 &  14151.152 & 	29.1 &    $-$25.511 &  14248.297 \\ 
 9.2 &  $-$3416.598 &  15539.253 & 	16.0 &   $-$321.074 &  13632.253 & 	22.6 &    $-$72.333 &  14153.993 & 	29.2 &    $-$25.155 &  14248.980 \\ 
 9.3 &  $-$3323.228 &  15403.620 & 	16.1 &   $-$312.077 &  13648.080 & 	22.7 &    $-$71.016 &  14156.762 & 	29.3 &    $-$24.806 &  14249.651 \\ 
 9.5 &  $-$3135.373 &  15140.684 & 	16.2 &   $-$303.409 &  13663.653 & 	22.8 &    $-$69.729 &  14159.462 & 	29.4 &    $-$24.462 &  14250.310 \\ 
 9.6 &  $-$3041.559 &  15014.060 & 	16.3 &   $-$295.056 &  13678.959 & 	22.9 &    $-$68.471 &  14162.094 & 	29.5 &    $-$24.124 &  14250.958 \\ 
 9.7 &  $-$2948.214 &  14891.067 & 	16.4 &   $-$287.003 &  13693.990 & 	23.0 &    $-$67.242 &  14164.660 & 	29.6 &    $-$23.793 &  14251.593 \\ 
 9.9 &  $-$2763.940 &  14656.966 & 	16.5 &   $-$279.238 &  13708.735 & 	23.1 &    $-$66.041 &  14167.163 & 	29.7 &    $-$23.466 &  14252.217 \\ 
10.0 &  $-$2673.450 &  14546.273 & 	16.6 &   $-$271.747 &  13723.190 & 	23.2 &    $-$64.867 &  14169.606 & 	29.8 &    $-$23.146 &  14252.831 \\ 
10.1 &  $-$2584.311 &  14440.039 & 	16.7 &   $-$264.519 &  13737.347 & 	23.3 &    $-$63.719 &  14171.988 & 	29.9 &    $-$22.831 &  14253.433 \\ 
10.2 &  $-$2496.684 &  14338.388 & 	16.8 &   $-$257.543 &  13751.204 & 	23.4 &    $-$62.597 &  14174.312 & 	30.0 &    $-$22.521 &  14254.025 \\ 
10.3 &  $-$2410.710 &  14241.415 & 	16.9 &   $-$250.806 &  13764.757 & 	23.5 &    $-$61.499 &  14176.580 & 	32.5 &    $-$16.235 &  14265.948 \\ 
10.4 &  $-$2326.513 &  14149.188 & 	17.0 &   $-$244.299 &  13778.004 & 	23.6 &    $-$60.426 &  14178.793 & 	35.0 &    $-$11.993 &  14273.903 \\ 
10.5 &  $-$2244.197 &  14061.746 & 	17.1 &   $-$238.013 &  13790.946 & 	23.7 &    $-$59.376 &  14180.953 & 	37.5 &     $-$9.045 &  14279.389 \\ 
10.6 &  $-$2163.851 &  13979.103 & 	17.2 &   $-$231.937 &  13803.580 & 	23.8 &    $-$58.349 &  14183.062 & 	40.0 &     $-$6.943 &  14283.276 \\ 
10.7 &  $-$2085.548 &  13901.253 & 	17.3 &   $-$226.064 &  13815.909 & 	23.9 &    $-$57.345 &  14185.121 & 	45.0 &     $-$4.275 &  14288.183 \\ 
10.8 &  $-$2009.347 &  13828.166 & 	17.4 &   $-$220.384 &  13827.933 & 	24.0 &    $-$56.362 &  14187.132 & 	50.0 &     $-$2.758 &  14290.958 \\ 
10.9 &  $-$1935.294 &  13759.794 & 	17.5 &   $-$214.890 &  13839.657 & 	24.1 &    $-$55.401 &  14189.096 & 	\\
\hline
\end{tabular}
\end{table*}

\begin{table*} 
\caption{Part I: Electric transition dipole moment between the X$^1\Sigma^+$ and A$^1\Sigma^+$ states of KCa$^+$.}
\label{tab:three}
\begin{tabular}{rr|rr|rr|rr}
\hline
        $R/a_0$  &   $d/(ea_0)$  &
        $R/a_0$  &   $d/(ea_0)$  &
        $R/a_0$  &   $d/(ea_0)$  &
        $R/a_0$  &   $d/(ea_0)$  
        \\
\hline
 8.0 &   2.219570 & 	14.0 &   1.540030 & 	19.6 &   0.234320 & 	25.2 &   0.021189 \\ 
 8.1 &   2.293330 & 	14.1 &   1.502300 & 	19.7 &   0.225129 & 	25.3 &   0.020253 \\ 
 8.2 &   2.358300 & 	14.2 &   1.464860 & 	19.8 &   0.216268 & 	25.4 &   0.019357 \\ 
 8.4 &   2.463930 & 	14.3 &   1.427760 & 	19.9 &   0.207726 & 	25.5 &   0.018500 \\ 
 8.6 &   2.541950 & 	14.4 &   1.391030 & 	20.0 &   0.199494 & 	25.6 &   0.017679 \\ 
 8.7 &   2.572480 & 	14.5 &   1.354690 & 	20.1 &   0.191563 & 	25.7 &   0.016894 \\ 
 8.8 &   2.598270 & 	14.6 &   1.318770 & 	20.2 &   0.183921 & 	25.8 &   0.016143 \\ 
 8.9 &   2.619900 & 	14.7 &   1.283300 & 	20.3 &   0.176563 & 	25.9 &   0.015424 \\ 
 9.0 &   2.637890 & 	14.8 &   1.248290 & 	20.4 &   0.169477 & 	26.0 &   0.014736 \\ 
 9.1 &   2.652670 & 	14.9 &   1.213780 & 	20.5 &   0.162656 & 	26.1 &   0.014078 \\ 
 9.2 &   2.664580 & 	15.0 &   1.179790 & 	20.6 &   0.156091 & 	26.2 &   0.013449 \\ 
 9.3 &   2.673930 & 	15.1 &   1.146320 & 	20.7 &   0.149772 & 	26.3 &   0.012847 \\ 
 9.5 &   2.685810 & 	15.2 &   1.113400 & 	20.8 &   0.143692 & 	26.4 &   0.012271 \\ 
 9.6 &   2.688690 & 	15.3 &   1.081030 & 	20.9 &   0.137844 & 	26.5 &   0.011721 \\ 
 9.7 &   2.689680 & 	15.4 &   1.049240 & 	21.0 &   0.132218 & 	26.6 &   0.011194 \\ 
 9.9 &   2.686360 & 	15.5 &   1.018030 & 	21.1 &   0.126808 & 	26.7 &   0.010690 \\ 
10.0 &   2.682160 & 	15.6 &   0.987412 & 	21.2 &   0.121607 & 	26.8 &   0.010208 \\ 
10.1 &   2.676320 & 	15.7 &   0.957396 & 	21.3 &   0.116606 & 	26.9 &   0.009748 \\ 
10.2 &   2.668860 & 	15.8 &   0.927988 & 	21.4 &   0.111799 & 	27.0 &   0.009307 \\ 
10.3 &   2.659810 & 	15.9 &   0.899192 & 	21.5 &   0.107179 & 	27.1 &   0.008886 \\ 
10.4 &   2.649170 & 	16.0 &   0.871014 & 	21.6 &   0.102741 & 	27.2 &   0.008484 \\ 
10.5 &   2.636980 & 	16.1 &   0.843456 & 	21.7 &   0.098476 & 	27.3 &   0.008099 \\ 
10.6 &   2.623240 & 	16.2 &   0.816520 & 	21.8 &   0.094378 & 	27.4 &   0.007732 \\ 
10.7 &   2.607960 & 	16.3 &   0.790206 & 	21.9 &   0.090443 & 	27.5 &   0.007380 \\ 
10.8 &   2.591170 & 	16.4 &   0.764515 & 	22.0 &   0.086664 & 	27.6 &   0.007045 \\ 
10.9 &   2.572890 & 	16.5 &   0.739446 & 	22.1 &   0.083034 & 	27.7 &   0.006723 \\ 
11.0 &   2.553140 & 	16.6 &   0.714995 & 	22.2 &   0.079550 & 	27.8 &   0.006417 \\ 
11.1 &   2.531960 & 	16.7 &   0.691161 & 	22.3 &   0.076204 & 	27.9 &   0.006124 \\ 
11.2 &   2.509370 & 	16.8 &   0.667939 & 	22.4 &   0.072993 & 	28.0 &   0.005844 \\ 
11.3 &   2.485420 & 	16.9 &   0.645324 & 	22.5 &   0.069912 & 	28.1 &   0.005576 \\ 
11.4 &   2.460150 & 	17.0 &   0.623313 & 	22.6 &   0.066954 & 	28.2 &   0.005320 \\ 
11.5 &   2.433610 & 	17.1 &   0.601900 & 	22.7 &   0.064116 & 	28.3 &   0.005076 \\ 
11.6 &   2.405850 & 	17.2 &   0.581070 & 	22.8 &   0.061394 & 	28.4 &   0.004843 \\ 
11.7 &   2.376930 & 	17.3 &   0.560829 & 	22.9 &   0.058782 & 	28.5 &   0.004620 \\ 
11.8 &   2.346900 & 	17.4 &   0.541157 & 	23.0 &   0.056277 & 	28.6 &   0.004407 \\ 
11.9 &   2.315830 & 	17.5 &   0.522055 & 	23.1 &   0.053874 & 	28.7 &   0.004204 \\ 
12.0 &   2.283790 & 	17.6 &   0.503510 & 	23.2 &   0.051569 & 	28.8 &   0.004010 \\ 
12.1 &   2.250830 & 	17.7 &   0.485514 & 	23.3 &   0.049361 & 	28.9 &   0.003824 \\ 
12.2 &   2.217030 & 	17.8 &   0.468058 & 	23.4 &   0.047242 & 	29.0 &   0.003647 \\ 
12.3 &   2.182450 & 	17.9 &   0.451132 & 	23.5 &   0.045212 & 	29.1 &   0.003478 \\ 
12.4 &   2.147170 & 	18.0 &   0.434726 & 	23.6 &   0.043265 & 	29.2 &   0.003317 \\ 
12.5 &   2.111250 & 	18.1 &   0.418830 & 	23.7 &   0.041399 & 	29.3 &   0.003162 \\ 
12.6 &   2.074770 & 	18.2 &   0.403434 & 	23.8 &   0.039610 & 	29.4 &   0.003015 \\ 
12.7 &   2.037790 & 	18.3 &   0.388526 & 	23.9 &   0.037896 & 	29.5 &   0.002875 \\ 
12.8 &   2.000380 & 	18.4 &   0.374097 & 	24.0 &   0.036254 & 	29.6 &   0.002741 \\ 
12.9 &   1.962610 & 	18.5 &   0.360135 & 	24.1 &   0.034680 & 	29.7 &   0.002613 \\ 
13.0 &   1.924540 & 	18.6 &   0.346630 & 	24.2 &   0.033172 & 	29.8 &   0.002490 \\ 
\hline
\end{tabular}
\end{table*}

\begin{table*} 
\caption{Part II: Electric transition dipole moment between the X$^1\Sigma^+$ and A$^1\Sigma^+$ states of KCa$^+$.}
\label{tab:four}
\begin{tabular}{rr|rr|rr|rr}
\hline
        $R/a_0$  &   $d/(ea_0)$  &
        $R/a_0$  &   $d/(ea_0)$  &
        $R/a_0$  &   $d/(ea_0)$  &
        $R/a_0$  &   $d/(ea_0)$  
        \\
\hline
13.1 &   1.886230 & 	18.7 &   0.333571 & 	24.3 &   0.031727 & 	29.9 &   0.002374 \\ 
13.2 &   1.847740 & 	18.8 &   0.320947 & 	24.4 &   0.030344 & 	30.0 &   0.002262 \\ 
13.3 &   1.809140 & 	18.9 &   0.308747 & 	24.5 &   0.029018 & 	32.5 &   0.000668 \\ 
13.4 &   1.770470 & 	19.0 &   0.296960 & 	24.6 &   0.027749 & 	35.0 &   0.000190 \\ 
13.5 &   1.731800 & 	19.1 &   0.285576 & 	24.7 &   0.026534 & 	37.5 &   0.000054 \\ 
13.6 &   1.693170 & 	19.2 &   0.274584 & 	24.8 &   0.025370 & 	40.0 &   0.000017 \\ 
13.7 &   1.654630 & 	19.3 &   0.263972 & 	24.9 &   0.024255 & 	45.0 &   0.000005 \\ 
13.8 &   1.616230 & 	19.4 &   0.253732 & 	25.0 &   0.023188 & 	50.0 &   0.000005 \\ 
13.9 &   1.578020 & 	19.5 &   0.243851 & 	25.1 &   0.022167 & 	\\
\hline
\end{tabular}
\end{table*}

\bibliography{KCaplus_refs}